\documentclass[aps,twocolumn,superscriptaddress,amsfont,graphicx,nofootinbib,preprintnumbers]{revtex4-1}%
\UseRawInputEncoding
\usepackage{amsmath, amsthm, amssymb}
\usepackage{graphicx}
\usepackage{subfigure}
\usepackage{wrapfig}
\usepackage{color}
\usepackage{color,graphicx,epsfig}
\usepackage{ifpdf}
\usepackage{bm}
\usepackage[english]{babel}
\usepackage{braket}
\usepackage{hyperref}
\usepackage{enumerate}
\usepackage{url}
\usepackage{multirow}
\usepackage{tablefootnote} 
\usepackage{cases}

\usepackage{slashed}

\usepackage{changes}
\begin{document}
\title{Probing Gauged $U(1)$ Sub-GeV Dark Matter via Cosmic Ray Cooling in Active Galactic Nuclei} 

\author{
	\href{https://orcid.org/0000-0001-8158-6602}{Arvind Kumar Mishra}
	${\href{https://orcid.org/0000-0001-8158-6602}{\includegraphics[height=0.15in,width=0.15in]{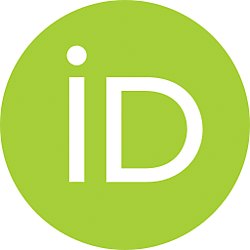}}}$ 
}
\email{arvindm215@gmail.com}
\affiliation{Department of Physics and Institute of Theoretical Physics, Nanjing Normal University, Nanjing, 210023, China}
\affiliation{Nanjing Key Laboratory of Particle Physics and Astrophysics, Nanjing, 210023, China}

\author{Ning Liu}
\email{liuning@njnu.edu.cn}
\affiliation{Department of Physics and Institute of Theoretical Physics, Nanjing Normal University, Nanjing, 210023, China}
\affiliation{Nanjing Key Laboratory of Particle Physics and Astrophysics, Nanjing, 210023, China}

\author{Chih-Ting Lu}
\email{ctlu@njnu.edu.cn}
\affiliation{Department of Physics and Institute of Theoretical Physics, Nanjing Normal University, Nanjing, 210023, China}
\affiliation{Nanjing Key Laboratory of Particle Physics and Astrophysics, Nanjing, 210023, China}
        
\date{\today}

\begin{abstract}
Cosmic rays (CRs) traversing the dark matter (DM) spike surrounding active galactic nuclei (AGNs) can be cooled through interactions with DM particles. In this study, we investigated constraints on sub-GeV DM particles charged under various $U(1)$ gauge symmetries by exploiting the cooling effect of CRs in AGNs. We find that for low DM and mediator masses, the CR cooling rate is higher compared to the standard model cooling process. Furthermore, by utilizing constraints from the CR cooling effect in NGC 1068 and TXS 0506+056, we explore the bounds on the DM-electron and DM-proton elastic scattering cross-sections. Our results indicate that in the sub-GeV DM mass range, these constraints are more stringent than those from certain boosted DM mechanisms and current direct detection limits.
\end{abstract}
		

		\maketitle

\section{Introduction}
The existence of dark matter (DM) has been confirmed on various cosmological scales, from galaxies to clusters, yet its particle nature remains unknown~\cite{Bertone:2004pz}. Current experiments, which are primarily sensitive to DM masses above the GeV scale, utilize direct detection, indirect detection, and collider searches, as reviewed in Ref.~\cite{Battaglieri:2017aum}. However, the non-detection of DM particles in these mass ranges has resulted in stringent constraints on the available parameter space \cite{Essig:2012yx,Aprile:2012zx,PandaX-II:2017hlx,Misiaszek:2023sxe}.

Exploring sub-GeV DM particles is a promising direction, as these light DM particles are challenging to probe in current direct detection experiments due to the low momentum transfer to detector materials~\cite{Essig:2022dfa,Essig:2011nj,Knapen:2017xzo,Lin:2022hnt,Cirelli:2023tnx,Balan:2024cmq}. Various detection techniques have been proposed to search for light DM particles, including DM-electron scattering~\cite{Essig:2011nj,Emken:2019tni}, DM absorption on electrons~\cite{Dror:2019dib}, DM-nucleus scattering via the Migdal effect~\cite{Ibe:2017yqa}, and bremsstrahlung~\cite{Kouvaris:2016afs}.   
In addition, novel methods to boost light DM particles offer a new avenue for probing DM, as they can transfer sufficient momentum to materials to overcome the detection thresholds in direct detection experiments and enhance the likelihood of detection \cite{Agashe:2014yua}. Previous studies have shown that employing these mechanisms to boost light DM particles can impose strong constraints on the DM parameter space~\cite{Agashe:2014yua,Giudice:2017zke,Cao:2020bwd,Bardhan:2022bdg,Bhowmick:2022zkj,Lin:2024vzy}.

The cooling of cosmic rays (CRs) in active galactic nuclei (AGNs) offers a novel mechanism to probe sub-GeV DM particles. As CRs traverse the DM spike surrounding AGNs, interactions with DM particles cool them down, thereby altering the high-energy neutrino and gamma-ray fluxes observed from these sources. By leveraging this CR-DM cooling effect and comparing the results with observational data, constraints on light DM particles can be derived. This approach was first applied in the case of a scalar mediator~\cite{Herrera:2023nww} and subsequently extended to the inelastic DM scenario~\cite{Gustafson:2024aom}. The promising constraints provided by this method encourage further exploration of other DM models to expand the scope of previous studies. Moreover, unlike direct detection experiments for cosmic ray-boosted dark matter, which often suffer from multiple scatterings between DM and SM particles~\cite{PandaX-II:2021kai,CDEX:2022fig,Super-Kamiokande:2022ncz,LZ:2025iaw}, the cooling effect requires only a single CR-DM interaction to trigger detection. This makes it a more efficient and complementary approach for probing light DM.

In this work, we explore the sub-GeV DM particles by examining the cooling of CRs in AGNs via the CR-DM collisions. In particular, we focus on various DM models with extra $U(1)$ gauge symmetries, including secluded dark sector models~\cite{Chun:2010ve,Alenezi:2025kwl}, $ U(1)_{B-L}$ models~\cite{Okada:2010wd,Basak:2013cga}, and $ U(1)_{L_e - L_\mu}$ models~\cite{Duan:2017qwj,Cao:2017sju}. We estimate the cooling time scale resulting from CRs-DM interactions and compare it with the predictions for cooling effects from standard model (SM) processes. We find that for both low DM masses and low dark photon masses, the cooling time is shorter than the predicted SM processes. Consequently, the constraints from cooling effects become more stringent when both the DM and the mediator are light, thereby extending to cover previously unconstrained regions of parameter space.

 By leveraging the CR cooling effect in AGNs, we constrain both the DM-electron and DM-proton scattering cross-sections. For the heavy mediator case, we find that electron-induced cooling imposes stringent constraints on low DM masses (i.e., $m_{\chi}\leq 2\times 10^{-4}$ GeV), which are more restrictive than those from current direct detection experiments~\cite{SENSEI:2020dpa,DarkSide:2022knj} and the solar reflection mechanism~\cite{An:2017ojc}. In contrast, for the light mediator case, the constraints are weaker and primarily limited by existing astrophysical bounds. Additionally, our analysis reveals that the CR cooling effect provides a strong constraint on the DM-proton elastic scattering cross-section, particularly for $m_{\chi}\leq 10^{-2}$ GeV, exceeding the limits from current direct detection~\cite{CRESST:2019jnq,SENSEI:2020dpa,Emken:2019tni} and the blazer-boosted DM mechanism~\cite{Wang:2021jic,CDEX:2024qzq}.

The arrangement of our work is as follows: In Section~\ref{sec:CRcool}, we briefly discuss the CR cooling in AGNs due to CRs-DM scattering. Further, we describe the three relevant gauged $U(1)$ DM models in Section~\ref{sec:models}. Then, we discuss our main results in Section~\ref{sec:results}. Finally, we summarize our findings in Section~\ref{sec:concl}.

\section{\label{sec:cooling} Cosmic ray cooling via CR-DM interactions}
\label{sec:CRcool}
Observations suggest that a supermassive black hole (SMBH) resides near the center of an AGN, serving as the primary engine for the energetic phenomena observed within these systems. CR electrons and protons passing near the SMBH can be accelerated, and the accelerated protons, in particular, can produce high-energy neutrinos and electromagnetic (EM) signatures via \(pp\) and \(p\gamma\) processes~\cite{Murase:2022feu}. Multimessenger observations of the AGNs NGC 1068 and TXS 0506+056 enable us to estimate the energy-dependent cooling times of CR protons and electrons. In NGC 1068, CR protons lose energy through Standard Model (SM) processes such as \(pp\) interactions, \(p\gamma\) interactions, and Bethe-Heitler pair production~\cite{Murase:2019vdl,Murase:2022dog}. Additionally, CR electrons in TXS 0506+056 are cooled by synchrotron radiation, inverse Compton scattering \cite{Herrera:2023nww}.

Furthermore, it has been argued that the adiabatic growth of an SMBH can lead to the formation of a DM spike in its vicinity. The resulting DM density profile depends on the intrinsic properties of the DM particles; for annihilating DM scenarios, such as those involving weakly interacting massive particles or self-interacting dark DM, the peak density is reduced due to annihilation~\cite{Chiang:2019zjj,Alvarez:2020fyo}. As CRs pass through the DM spike created by the SMBH's gravitational field, they scatter off DM particles, transferring energy to them. If the cooling rate from CR-DM scattering exceeds that predicted by SM processes, then CR cooling effects become more pronounced, a scenario that is not achievable under standard conditions. Therefore, a significant CR cooling effect is expected whenever a DM spike is present.

The cooling timescale due to elastic DM-CRs scattering is given by \cite{Ambrosone:2022mvk}
\begin{equation}\label{eq:tau_el}
    \tau^{\mathrm{cool}}_{\chi-i}=\Bigg[-\frac{1}{E}\bigg(\frac{dE}{dt}\bigg)_{\chi i}\Bigg]^{-1},
\end{equation}
where $E$ is the energy of CRs and $dE/dt$ is the energy loss rate during CR and DM collisions. The  energy loss rate of CRs is given by
\begin{equation}\label{eq:dmenergyloss}
    \left(\frac{dE}{d t}\right)_{\chi i} = -\frac{\langle \rho_{\chi}\rangle}{{m_{\chi}}}\,\int_{0}^{T^{\rm max}_{\chi}} dT_{\chi}\, T_{\chi} ~\frac{d\sigma_{\chi i}}{dT_{\chi}} \,,
\end{equation}
where $m_{\chi}$, and $\langle \rho_{\chi} \rangle$ represent mass and the average density of DM particles in the region of CR production, and $i=e,\ p$ represents the constituents of the CRs.  Further, $d\sigma_{\chi i}/dT_{\chi}$ represents the differential elastic cross-section between DM-proton (or DM-electron) in which $T_{\chi}$ is the final kinetic energy of the DM particle. Further, $T^{\rm max}_{\chi}$ is the maximum kinetic energy value of DM particles resulting during DM-CRs collision, and given by
\begin{equation}\label{eq:energygain}
    T^{\rm max}_{\chi} = \frac{2T^2 + 4m_i T}{m_{\chi}}\left[\left(1+\frac{m_i}{m_{\chi}}\right)^2 + \frac{2 T}{m_{\chi}}\right]^{-1} \,
\end{equation}
where $m_i$ and  $T$ represent the mass and kinetic energy of the particles that constitute the CR.  Furthermore, it is crucial to mention that during the DM scattering with the protons, whenever $E\geq m^{2}_{p}/2m_{\chi}$, inelastic scattering will dominate over the elastic case. Therefore, we will restrict our analysis and work within the elastic limit.  

\section{\label{sec:models} Gauged $U(1)$ models for CR-DM interactions} 
\label{sec:models}
Now, we consider three typical gauged $U(1)$ models of the CR-DM collisions. In this work, we extend the SM gauge group by incorporating an additional $U(1)$ gauge symmetry to obtain these three DM models. 
Here, secluded dark sector, $ U(1)_{B-L}$, and $ U(1)_{L_e - L_\mu}$ models with Dirac fermionic DM are considered. In these models, the common Lagrangian terms can be written as follows: 
 \begin{align}
   \mathcal{L} = & \mathcal{L}_{SM} + \bar{\chi} \left(i \slashed{\partial} - m_\chi \right) \chi +\frac{1}{2} m_{A'}^2 A'_\mu A'^{\mu} +\mathcal{L}_{int},
  \label{eq:LagsSecl}
 \end{align}
where $\mathcal{L}_{SM}$ represents the SM Lagrangian, $m_\chi$ and $m_{A'}$ are DM and dark photon masses. We represent $\mathcal{L}_{int} = \mathcal{L}_{Sec}$ for the secluded dark sector model, $\mathcal{L}_{int} = \mathcal{L}_{B-L}$ for the $U(1)_{B-L}$ model, and $\mathcal{L}_{int} = \mathcal{L}_{L_e-L_\mu}$ for the $U(1)_{L_e-L_\mu}$ model, respectively. These interaction terms are represented below: 
\begin{enumerate}
    \item 
Secluded dark sector: 
 \begin{align}
   \mathcal{L}_{Sec} = & {A'}_\mu \left[ g_{fL}^{A'}~\bar{f} \gamma^\mu P_L f + g_{fR}^{A'}~\bar{f} \gamma^\mu P_R f + g_{\chi}^{A'}~\bar{\chi} \gamma^\mu \chi \right] \nonumber \\
 &+ {Z}_\mu \left[ g_{fL}^{Z}~\bar{f} \gamma^\mu P_L f + g_{fR}^{Z}~\bar{f} \gamma^\mu P_R f + g_{\chi}^{Z}~\bar{\chi} \gamma^\mu \chi \right]\,
  \label{eq:LagsSecl}
 \end{align}
where $P_{L,R}$ represent the chiral projection operators, defined as $P_{L,R}=\frac{1}{2}(1\mp\gamma_5)$  and $f$ is the SM fermions. The detailed forms of the couplings \(g_{fL}^{A'}\), \(g_{fR}^{A'}\), \(g_{\chi}^{A'}\), \(g_{fL}^{Z}\), \(g_{fR}^{Z}\), and \(g_{\chi}^{Z}\) can be found in the Appendix of Ref.~\cite{Chun:2010ve}. Moreover, due to the kinetic mixing between the new \(U(1)\) and the SM \(U(1)_Y\) gauge symmetries, we denote the kinetic mixing parameter by \(\epsilon\). Consequently, all these couplings are functions of \(\epsilon\). 

 \item $U(1)_{B-L}$: 
 \begin{equation}
  \mathcal{L}_{B-L} = g_{B-L} \left[- \bar{l} \gamma^\mu A^{\prime}_{\mu} l + \frac{1}{3} \bar{q} \gamma^\mu A^{\prime}_{\mu} q \right] 
 - g_{\chi} \bar{\chi} \gamma^\mu \chi {A'}_\mu\, ,
  \label{eq:LagBL}
 \end{equation} 
where $l$ and $q$ are SM leptons and quarks.  

\item 
$U(1)_{L_{e} - L_{\mu}}$: 
\begin{equation}
 \mathcal{L}_{L_{e} - L_{\mu}} = g_{L} \left[\bar{l}_e \gamma^\mu A'_{\mu} l_e -\bar{l}_\mu \gamma^\mu A'_{\mu} l_\mu  \right]- g_{\chi} \bar{\chi} \gamma^\mu \chi {A'}_\mu\,, 
  \label{eq:LagLeLmu}
\end{equation} 
where $l_e = e, \nu_e$ and $l_\mu = \mu, \nu_\mu$. 
\end{enumerate}

It is clear that the dark photon couples to DM particles via a vector interaction. However, its interactions with SM fermions differ among these three models. In the secluded dark sector model, the dark photon exhibits both vector and axial-vector couplings to all SM fermions. In the \(U(1)_{B-L}\) model, it couples only via vector interactions to pairs of leptons and quarks, and in the \(U(1)_{L_{e}-L_{\mu}}\) model, it couples solely via vector interactions to pairs of leptons. Consequently, the constraints from the CR cooling effect will differ for these models.

The scattering amplitude and differential cross-section rate of CR-DM scattering for these models are given by~\cite{Cao:2020bwd,Guha:2024mjr}
\footnote{In our analysis, we consider the case of a light mediator, i.e., \( m_{A'} \ll m_{Z} \), where \( m_{Z} \) is the $Z$ boson mass. Consequently, in the secluded dark sector, the contribution from $Z$ boson mediator will be subdominant compared to that from the dark photon. Therefore, this formulation remains general and applicable to the three DM models considered in Ref.~\cite{Cho:2020mnc}. For a complete cross-section calculation, including contributions from $Z$ boson mediator, we refer to Refs.~\cite{Cho:2020mnc,Guha:2024mjr}.}
\begin{align}
& \overline{{\left|\mathcal{M}_{\chi i}\right|}^2 } = {g_{i}^{A'}}^2 {g_{\chi}^{A'}}^2 
 \nonumber \\ 
  &\times \frac{2m_\chi \left(m_i + T_i \right)^2 - T_\chi \left\lbrace \left( m_i + m_\chi \right)^2 + 2 m_\chi T_i  \right\rbrace + m_\chi T_\chi^2}{ \left(2 m_\chi T_\chi + m_{A'}^2 \right)^2},
 \end{align}
  \begin{align}
& \frac{d \sigma_{\chi i}}{dT_\chi} = {g_{i}^{A'}}^2 {g_{\chi}^{A'}}^2 F^{2}_{i}\left(q^2\right)  
 \nonumber \\ 
  &\times \frac{2m_\chi \left(m_i + T_i \right)^2 - T_\chi \left\lbrace \left( m_i + m_\chi \right)^2 + 2 m_\chi T_i  \right\rbrace + m_\chi T_\chi^2}{4 \pi \left(2 m_i T_i + T_i^2 \right) \left(2 m_\chi T_\chi + m_{A'}^2 \right)^2},
 \label{eq:gencs}
 \end{align} 
where, $i=e,p$ represents the constituents of the CRs, and $q$ is the momentum transfer. Further, $F\left(q^2\right)$ is a form factor for the hadronic elastic scattering and given by ~\cite{Bringmann:2018cvk,Perdrisat:2006hj},
\begin{equation}
 F\left(q^2\right) = \frac{\Lambda^2}{q^{2}+\Lambda^2},\,
\end{equation}
 where $\Lambda= 770~\rm{MeV}$ \cite{Perdrisat:2006hj}. For a pointlike structure like electrons  $F\left(q^2\right)\rightarrow 0$. Further, in all the models, the couplings will be different, and for the $L_{e} - L_{\mu}$ model, DM will only interact with the electrons. Furthermore, in these models, the coupling forms differ, and the couplings in Eq.~(\ref{eq:gencs}) are given as follows: 
 \begin{itemize}
 \item Secluded dark sector model: 
$g_{e}^{A'} = g_{p}^{A'} = e\varepsilon\cos\theta_W$, and $g_{\chi}^{A'} = g_{\chi}$, with $\theta_W $ being the Weinberg angle; 
 \item 
The $ U(1)_{B-L}$ model: $g_{e}^{A'} = g_{p}^{A'}=  g_{B-L}$, and $g_{\chi}^{A'} = g_{\chi}$; 
 \item 
The $U(1)_{L_e - L_\mu}$ model: $g_{e}^{A'} =  g_{L}$, and $g_{\chi}^{A'} = g_{\chi}$. 
 \end{itemize}

In addition, to facilitate comparison with the existing experiments, we
define a reference cross-section for DM-electron scattering in a model-independent way as~\cite{Essig:2011nj}
\begin{equation}
\overline{\sigma}_e \equiv \frac{\mu_{\chi e}^2}{16 \pi m_\chi^2 m_e^2} \, \overline{\left|\mathcal{M}_{\chi i}(q_{\mathrm{ref}})\right|^2} \,
\label{eq:sigmaDMe}
\end{equation}
where $\mu_{\chi e}$ is the reduced DM-electron mass and $q_{\mathrm{ref}}=\alpha m_e$. Furthermore, in the DM-proton scattering case, we consider a non-relativistic cross-section in the assumption of heavy mediator as
\begin{equation}
\sigma_{\chi p}=\frac{{g_{i}^{A'}}^2 {g_{\chi}^{A'}}^2 \mu_{\chi p}^2}{\pi m^{4}_{A'}}
\label{eq:sigmaDMp}
\end{equation}
where $\mu_{\chi p}$ is the reduced DM-proton mass. Therefore, in Section~\ref{sec:results}, constraints will be shown in $\overline{\sigma}_{e}-m_{\chi}$ plane and $\sigma_{\chi p}-m_{\chi}$ plane for the DM-electron and DM-proton scattering cross-sections, respectively. 
\begin{figure*}
   \includegraphics[height=2in,width=3.in]{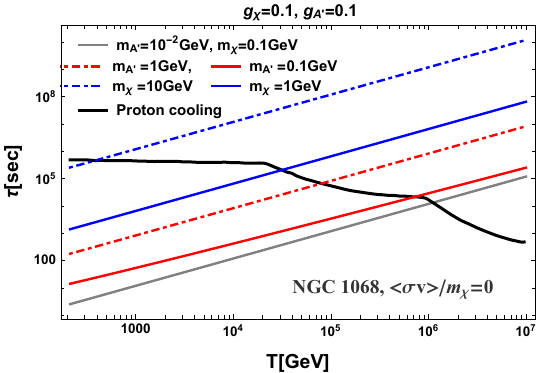}\hfil
 \includegraphics[height=2in,width=3.in]{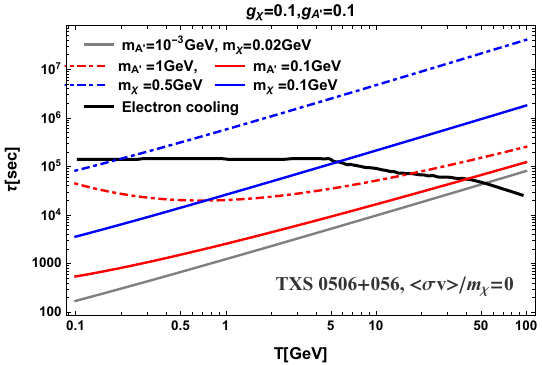}\par
\caption{Cooling timescale from the CR-DM elastic scattering with the kinetic energy of the CRs. The \textit{Left panel} and \textit{Right panel} correspond to the proton-DM and electon-DM scattering, respectively. For both panels, a solid black line is CR cooling through the standard model processes. Here, we fix the coupling constants $g_{\chi} = g_{A'} = 0.1$, while varying the masses $m_{\chi}$ and $m_{A'}$. } 
	\label{fig:coolingtime}%
\end{figure*}
\begin{table*}[t!]
		\begin{center}
			\begin{tabular}{l|ccccccc}
				\hline
				&  $\langle \sigma v \rangle/m_{\chi}$ & $\langle \rho_{\chi} \rangle$  \\
				\hline
				\toprule
				 NGC 1068 (A) &  0 & $5\times10^{18}$ GeV/cm$^{3}$   \\
				 NGC 1068 (B) & $10^{-31}$cm$^{3}$s$^{-1}$/GeV & $4\times10^{13}$ GeV/cm$^{3}$      \\
				 TXS 0506+056 (A) &   0 & $8\times10^{12}$ GeV/cm$^{3}$  \\
				 TXS 0506+056 (B) & $10^{-28}$cm$^{3}$s$^{-1}$/GeV & $4\times10^{11}$ GeV/cm$^{3}$   \\
                \hline
			\end{tabular}
		\end{center}
        \caption{The DM parameters, i.e., average DM density (within a certain radius), average DM-annihilation cross-section in NGC 1068 and TXS 0506+056 \cite{Herrera:2023nww}. Furthermore, A and B correspond to the non-annihilating and annihilating DM particles with the defined cross-section values. }
		\label{tab:data}
	\end{table*}
\begin{figure*}
   \includegraphics[height=2in,width=3in]{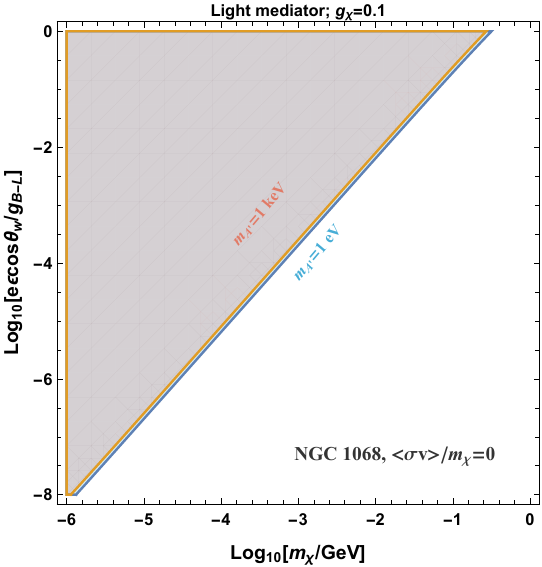}\hfil
 \includegraphics[height=2in,width=3in]{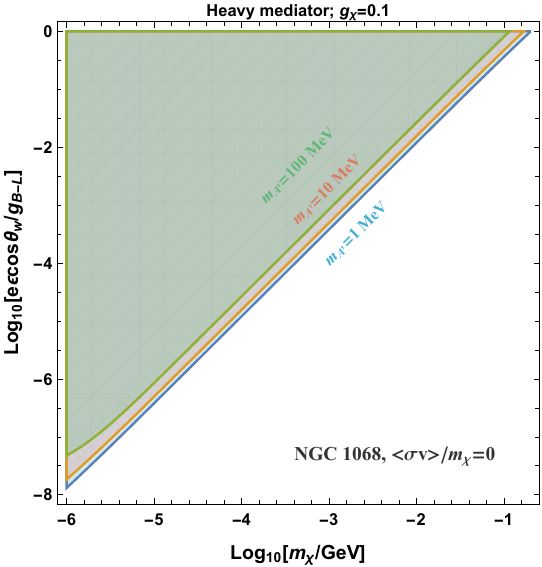}\par
\caption{Constraint on the DM-proton coupling as a function of DM mass from the proton cooling effect. \textit{Left panel} and \textit{Right panel} correspond to light and heavy mediators, respectively. The constraints are obtained using the NGC 1068 observations with non-annihilating (asymmetric) DM particles, i.e., $\langle \sigma v \rangle/m_{\chi}=0$. Here, we fix $g_{\chi}=0.1$, while varying the values of $m_{A'}$. Since DM couples to protons only in the secluded dark sector and \(U(1)_{B-L}\) models, we present the coupling constraints for these two models.}
	\label{fig:mediatorP}%
\end{figure*}
\section{Result and discussions}
\label{sec:results}
After being equipped with the necessary basic ingredients (the cooling mechanism and concrete DM models), we now show our results in the assumption of the elastic scattering between CRs and DM. To account for the cooling effects induced by SM processes, we focus on NGC 1068 and TXS 0506+056, for which multimessenger data and astrophysical models are available. These data allow us to estimate the cooling time as a function of CR energy (see Ref.~\cite{Herrera:2023nww} and references therein). Further, for a given SMBH parameters of  NGC 1068 and TXS 0506+056, the peak density profile of DM particles has also been calculated for both non-annihilating and annihilating DM particles \cite{Herrera:2023nww}. We summarize the average DM density for different DM scenarios in Table~\ref{tab:data}, see Ref.~\cite{Herrera:2023nww}, and references therein. It is evident that increasing the DM annihilation cross-section leads to a reduction in the DM density. In our estimation of CR cooling via CR-DM interactions, we will utilize the above DM parameters and cooling time scale (as mentioned latter in Fig.~\ref{fig:coolingtime}) unless otherwise defined explicitly. 

\subsection{CR cooling via elastic scattering with DM}
\label{subsec:cooling} 

Fig.~\ref{fig:coolingtime} shows the variation of the cooling timescale (due to elastic CRs-DM scattering) as a function of the CR's kinetic energy with fixed $g_{\chi} = g_{A'} = 0.1$. Here, \textit{Left panel} represents the cooling time due to proton-DM elastic scattering in the NGC 1068 for non-annihilating (asymmetric) DM particles, i.e., $\langle \sigma v \rangle/m_{\chi}=0$. The base gray line corresponds to the $m_{A'}=10^{-2}$ GeV and $m_{\chi}=0.1$ GeV. Further, the solid (dashed) red lines represent the cooling time due to enhanced mediator mass, whereas the solid (dashed) blue line corresponds to increasing DM mass. The solid black line is CR protons cooling through via the SM processes in NGC 1068 \cite{Murase:2022dog}. The cooling time increases with the kinetic energy of CRs. Also, the timescale strongly depends on the DM and mediator masses. We find that increasing the DM mass or mediator mass enhances the cooling time and vice versa. Therefore, cooling of CRs is mostly relevant for low masses of DM and mediator particles.

Further, \textit{Right panel} of Fig.~\ref{fig:coolingtime} represents the cooling time due to electron-DM elastic scattering for TXS 0506+056 non-annihilating (asymmetric) DM particles. The base gray line corresponds to the $m_{A'}=10^{-3}$ GeV and $m_{\chi}=2\times 10^{-2}$ GeV. Here, the red and blue lines represent the cooling time due to enhanced mediator and  DM masses. The solid black line is CR electrons cooling using the SM processes in TXS 0506+056 \cite{Keivani_2018}. 
We observe that increasing either the DM mass or the mediator mass results in a longer cooling timescale, as expected. Furthermore, a notable feature is that, at low DM masses, electron-DM scattering leads to lower energy transfer and consequently a longer cooling timescale compared to the proton case. 

\begin{figure*}
   \includegraphics[height=2in,width=3in]{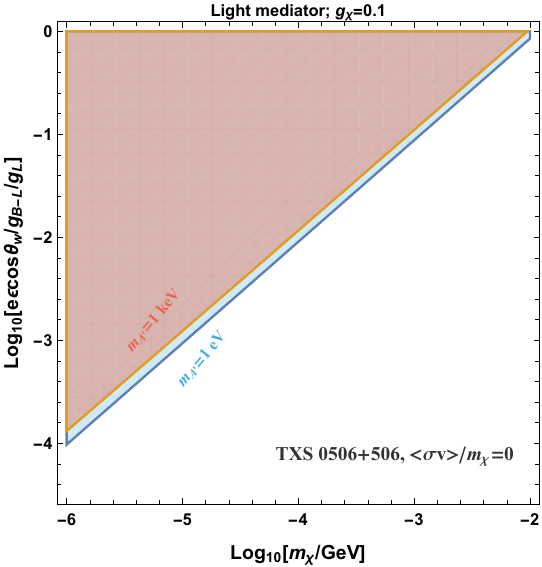}\hfil
 \includegraphics[height=2in,width=3in]{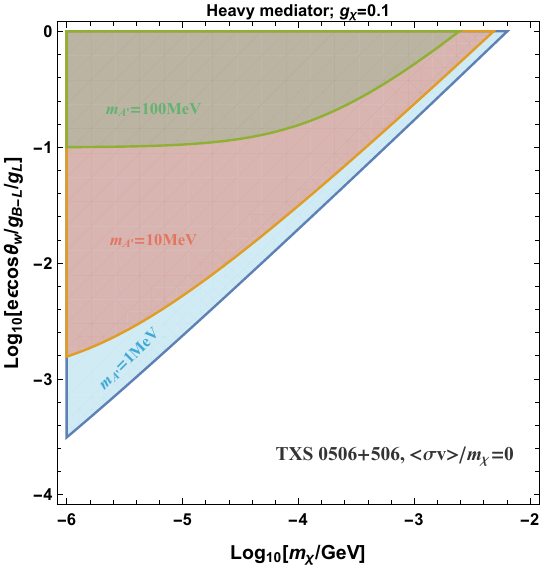}\par
\caption{Constraint on the DM-electron coupling as a function of DM mass from the electron cooling effect. \textit{Left panel} and \textit{Right panel} correspond to light and heavy mediators, respectively. The constraints are obtained using the TXS 0506+056 observations with non-annihilating (asymmetric) DM particles, i.e., $\langle \sigma v \rangle/m_{\chi}=0$. Here, we fix $g_{\chi}=0.1$, while varying the values of $m_{A'}$. Since DM couples to electrons in all \(U(1)\)-extended models, coupling constraints are provided for all the models under consideration.}
	\label{fig:gvsDMmassE}%
\end{figure*}
\subsection{Constraints on the DM microphysics}

Now, we apply the CR cooling condition to derive constraints on these three gauged 
$U(1)$ DM models. The CR cooling condition is given by:
\begin{equation}
\tau^{\mathrm{cool}}_{\chi-i}\leq C~\tau^{cool}_{\mathrm{SM}}~~,
\end{equation}
where $\tau^{cool}_{\mathrm{SM}}$ is the cooling timescale due to the SM processes as given in Fig.~\ref{fig:coolingtime}. Further, $C$ is model dependent, however, in our calculation, we will assume $C=0.1$ ~\cite{Herrera:2023nww}, otherwise stated explicitly.

Fig.~\ref{fig:mediatorP}  shows the coupling and DM mass parameter space for proton-DM scattering from CR cooling in  NGC 1068 for non-annihilating (asymmetric) DM particles, i.e., $\langle \sigma v \rangle/m_{\chi}=0$.  \textit{Left panel} and \textit{Right panel} correspond to the light and heavy mediator cases. In our considered DM models, interactions between DM particles and protons appear only in the secluded dark sector and \(U(1)_{B-L}\) models. Therefore, the constraints on couplings are presented only for these two models.
The blue and light red regions correspond to $m_{A'} = 1$ eV, and  $m_{A'}= 1$ keV, while keeping $g_{\chi}=0.1$. We observe that the proton cooling effect imposes strong constraints on the coupling for low DM masses; however, increasing the mediator mass from eV to keV does not significantly alter these constraints. This is because the CR cooling effect is more efficient for low DM and mediator masses, leading to stronger restrictions. In the heavy mediator case, while the coupling remains strongly constrained for low DM masses, increasing the dark photon mass from 1 MeV to 100 MeV results in a significant reduction of the allowed parameter space.

\begin{figure*}
   \includegraphics[height=2in,width=3in]{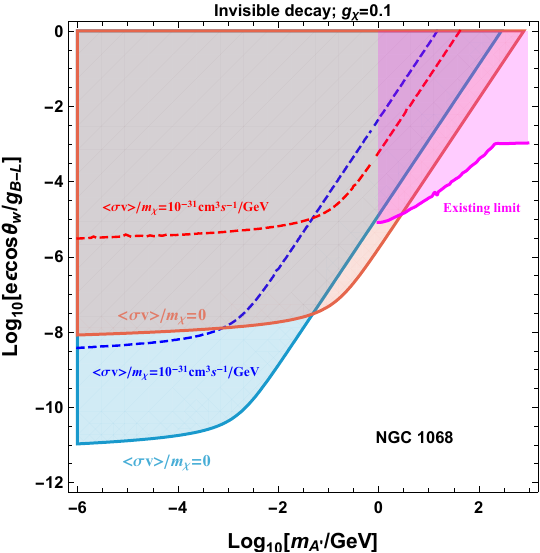}\hfil
 \includegraphics[height=2in,width=3in]{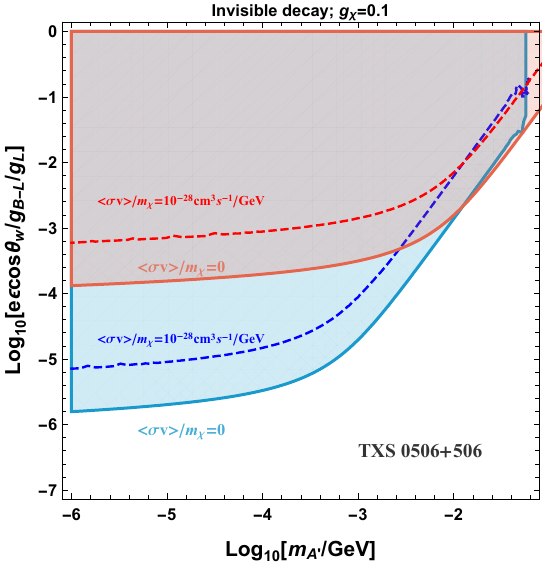}\par
\caption{Constraint on the DM-CRs coupling as a function of dark photon mass from the invisible decay. \textit{Left panel} and \textit{Right panel} correspond to DM-proton and DM-electron scattering, respectively. Note that solid blue (red) color regions represent $m_{\chi}=1$ MeV, and $m_{\chi}=100$ MeV using $\langle \sigma v \rangle/m_{\chi}=0$. However, the dashed blue (red) lines correspond to $m_{\chi}=1$ MeV, and $m_{\chi}=100$ MeV using annihilating DM particles with $\langle \sigma v \rangle/m_{\chi}=10^{-31}cm^{3}s^{-1}/$GeV. The existing limits are shown in magenta color~\cite{BaBar:2017tiz,Andreev:2021fzd,Essig:2013vha}.  Here, $g_{\chi}=0.1$ is fixed. }
	\label{fig:gvsDMmassP}%
\end{figure*}
\begin{figure*}
   \includegraphics[height=2in,width=3.2in]{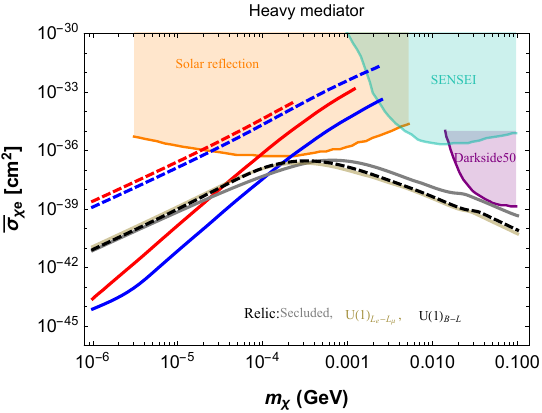}\hfil
 \includegraphics[height=2in,width=3.2in]{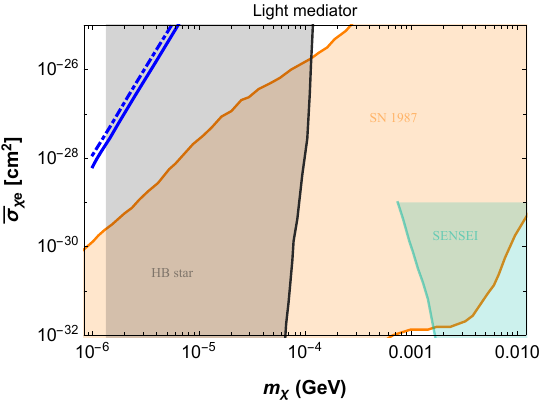}\par
\caption{ Constraint on the DM-electron elastic scattering cross-section from the CRs cooling effect. \textit{Left panel} and \textit{Right panel} represent the constraints from the heavy mediator and light mediator cases, respectively. For comparison, other existing constraints, including HB star \cite{Hardy:2016kme,Bardhan:2022bdg}, SN 1987 \cite{Chang:2018rso}, solar reflection \cite{An:2017ojc}, and SENSEI \cite{SENSEI:2020dpa} and Darkside-50 \cite{DarkSide:2022knj}, are also shown.
In addition, the relic density constraints are also shown for three gauged $U(1)$ DM models based on the freeze-out mechanism. Solid, dashed, and dot-dashed lines are explained in the main text. } 
\label{fig:cselectron}
\end{figure*}
Fig.~\ref{fig:gvsDMmassE} shows the coupling and DM mass parameter space for electron-DM scattering using CR cooling in TXS 0506+056 with non-annihilating DM particles. \textit{Left panel} and \textit{Right panel} correspond to the light and heavy mediator cases. Furthermore, unlike DM-proton scattering, DM particles interact with electrons in all \(U(1)\)-extended models. Therefore, coupling constraints are provided for all these models, including the secluded dark sector, \(U(1)_{B-L}\), and \(U(1)_{L_e - L_\mu}\) models. In the light mediator scenario, the coupling parameter is tightly constrained at low DM masses and shows only minor variation when the dark photon mass increases from the eV to keV scale. In contrast, in the heavy mediator case, increasing the dark photon mass significantly relaxes the constraints on the coupling parameter.

Now, as discussed, if the dark photon mass is higher than twice of the DM mass, i.e., $m_{A'}>2m_{\chi}$, then dark photons can decay into a pair of DM particles. Therefore, searching for an invisible decay of $A'$ will also constrain the parameter space. In Fig.~\ref{fig:gvsDMmassP}, we plot the parameter space of DM coupling with CRs and dark photon mass from the invisible decay of the dark photons using CRs cooling \footnote{We also explored the constraint on the DM model parameters from the visible decay of the dark photons, i.e., $m_{\chi}>m_{A'}/2$. We found that for the assumed model parameters, CR cooling will weakly constrain the DM parameter space. Therefore, we have not shown these constraints in our discussion.}. 
\textit{Left panel} of Fig.~\ref{fig:gvsDMmassP} corresponds to constraints on coupling and DM mass due to proton-DM scattering in NGC 1068. Here, solid blue (red) color regions represent $m_{\chi}=1$ MeV, and $m_{\chi}=100$ MeV using $\langle \sigma v \rangle/m_{\chi}=0$. Further, the dashed blue (red) lines correspond to $m_{\chi}=1$ MeV, and $m_{\chi}=100$ MeV using annihilating DM particles with $\langle \sigma v \rangle/m_{\chi}=10^{-31}cm^{3}s^{-1}/$GeV. The existing constraints are shown in magenta color and considered from Refs. \cite{BaBar:2017tiz,Andreev:2021fzd,Essig:2013vha}. It is clear that an increase in DM mass reduces the coupling strength, as discussed previously. Moreover, the presence of non-vanishing DM annihilation further relaxes the coupling strength. This is because annihilating DM reduces the DM density in the spike region, thereby allowing less energy to be transferred to DM. Consequently, the cooling timescale increases. The constraints on the coupling are extremely stringent, on the order of $\mathcal{O}(10^{-11})$, for dark photon masses in the range $10^{-6} \mathrm{GeV} \leq m_{A^{\prime}}\leq 10^{-3}$ GeV.  \textit{Right panel} of Fig.~\ref{fig:gvsDMmassP} shows constraints on coupling and DM mass due to electron-DM scattering in TXS 0506+056 for both annihilating and non-annihilating DM particles. Here, we see that the constraint on the annihilating DM is weaker and follows a similar behavior as the proton-DM scattering case. In this scenario, the constraints on coupling are strong, $\mathcal{O}(10^{-6})$ for $10^{-6} \mathrm{GeV} \leq m_{A^{\prime}}\leq 10^{-3}$ GeV.

\begin{figure*}
   \includegraphics[height=2in,width=3.2in]{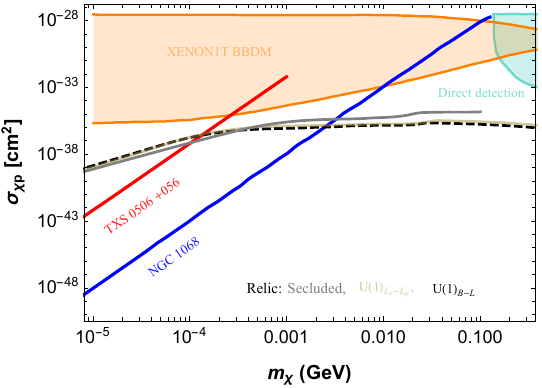}
\caption{ Constraint on the DM-proton elastic scattering cross-section from the CRs cooling effect in the heavy mediator case. For comparison, XENON1T BBDM \cite{Wang:2021jic,CDEX:2024qzq} and direct detection constraints \cite{CRESST:2019jnq,SENSEI:2020dpa,Emken:2019tni} are also included. In addition, the relic density constraints are also shown for three gauged $U(1)$ DM models based on the freeze-out mechanism. Here, the solid red and blue lines represent the constraints on the DM-proton cross-section derived from the CR cooling effect in TXS 0506+056 (A) and NGC 1068 (A), respectively, with \(m_{A'} = 100\) MeV fixed. } 
\label{fig:csproton}
\end{figure*}

After being equipped with the coupling constraint from CRs cooling, we will now also constrain DM-electron and DM-proton scattering cross-sections. To obtain the bound on DM-electron elastic scattering cross-section, we fix $g_{\chi}=0.1$ and use coupling constraint on both heavy and light mediators from Fig.~\ref{fig:gvsDMmassE}. Then we apply coupling constraint in Eq. (\ref{eq:sigmaDMe}), and estimate the constraint on reference cross-section from CRs cooling. Fig.~\ref{fig:cselectron} shows the constraint on the DM-electron elastic cross-section for three DM models with $U(1)$ gauge symmetry using the CR cooling constraint in TXS 0506+056. \textit{Left panel} displays the constraint on DM-electron cross-section in the heavy mediator case. The orange region represents the constraint from the solar reflection mechanism~\cite{An:2017ojc}. The green region represents the constraint from the direct detection experiment (SENSEI)~\cite{SENSEI:2020dpa}, and the purple region is from Darkside50~\cite{DarkSide:2022knj}. Further, we also estimate the relic abundance using the software \textit{micrOMEGAs}~\cite{Belanger:2010pz} for all \(U(1)\)-extended models in a freeze-out scenario. To do this, we fix \(g_{\chi} = 0.1\) and \(m_{A'} = 100\) MeV and then scan the dark photon coupling to SM particles as a function of the DM mass, ensuring consistency with the observed DM relic abundance, \(\Omega_{\text{DM}} h^2 = 0.1198 \pm 0.0015\)~\cite{Planck:2015ica}. We then use Eq.~(\ref{eq:sigmaDMe}) to estimate the constraint on the DM-electron scattering cross-section. The predicted relic abundances are illustrated by gray, brown, and black lines for the secluded dark sector, \(U(1)_{L_{e}-L_{\mu}}\), and \(U(1)_{B-L}\) models, respectively. Furthermore, the solid and dot-dashed blue lines are plotted for \(m_{A'} = 10\) MeV and \(100\) MeV, corresponding to TXS 0506+056 (A) (non-annihilating DM). Additionally, the constraints for TXS 0506+056 (B) (annihilating DM) are shown using the same mediator masses, represented by the solid and dot-dashed red lines.

Nevertheless, \textit{Left panel}  of Fig. \ref{fig:cselectron} indicates that the constraints for annihilating DM are weaker than those for non-annihilating DM, which is expected, as annihilation leads to a decrease in the DM density in the DM spike region. Furthermore, we find that in the case of a light mediator, the constraints on the cross-section are particularly stringent. This is because a light mediator (or light DM) leads to a faster cooling rate, which in turn results in stronger constraints compared to scenarios with heavier DM (see Section~\ref{subsec:cooling}). Although the constraints from the CR cooling effect can extend to relatively larger DM masses, those regions have already been excluded by the solar reflection mechanism and direct detection experiments. Therefore, we conclude that the electron cooling effect imposes stringent constraints on low DM masses, with \(m_{\chi} \leq 2 \times 10^{-4}\) GeV for TXS 0506+056 (A) and \(m_{\chi} \leq 2 \times 10^{-5}\) GeV for TXS 0506+056 (B).

\textit{Right panel} of Fig. \ref{fig:cselectron} represents the constraint of the light mediator using TXS 0506+056. 
Here, the black region represents a Horizontal Branch (HB) star's constraint \cite{Hardy:2016kme,Bardhan:2022bdg}, and the orange region corresponds to the constraint from the SN 1987 \cite{Chang:2018rso}. The green region represents the constraint from the direct detection experiment \cite{SENSEI:2020dpa}. The blue solid and dot-dashed lines represent the electron cooling constraint for $m_{A^{\prime}}=1$ eV, and $m_{A^{\prime}}=1$ keV. We find that for the light mediator case, the constraints are limited by the existing HB star constraint. Therefore, the electron cooling effect for the light mediator case does not provide strong constraints on DM-electron scattering.

Furthermore, to constrain DM-proton scattering, we follow the same procedure as for the DM-electron scattering case. After applying the coupling constraint for a heavy mediator from Fig.~\ref{fig:mediatorP} in Eq.~(\ref{eq:sigmaDMp}), we estimate the DM-proton elastic scattering cross-section from CR cooling while keeping \(g_{\chi} = 0.1\) fixed. Fig.~\ref{fig:csproton} shows the constraint on the DM-proton elastic cross-section for three gauged $U(1)$ DM models using the CR cooling constraint in AGNs. Here, we focus on the heavy mediator case. The orange region represents the constraint from the direct detection of Blazar-Boosted Dark Matter (BBDM) in XENON1T \cite{Wang:2021jic,CDEX:2024qzq}, and the green region is from a direct detection constraint \cite{CRESST:2019jnq,SENSEI:2020dpa,Emken:2019tni}. Further, the relic density constraints are shown based on the freeze-out mechanism. As discussed earlier, gray, brown, and black lines show the predicted relic abundance for the secluded dark sector, $U(1)_{L_{e}-L_{\mu}}$, and $U(1)_{B-L}$ models. 
Here, the solid red and blue lines represent the constraints on the DM-proton cross-section derived from the CR cooling effect in TXS 0506+056 (A) and NGC 1068 (A), respectively, with \(m_{A'} = 100\) MeV fixed. We find that in the low DM mass range, the CR cooling effect imposes strong constraints on the DM-proton elastic cross-section. Notably, the constraint from NGC 1068 is stronger than that from TXS 0506+056. Specifically, using NGC 1068 data, the bounds are more stringent for \(m_{\chi} \le 10^{-2}\) GeV, while with TXS 0506+056 (A) data, the constraints are stronger for \(m_{\chi} \le 2\times 10^{-4}\) GeV. 

\section{CONCLUSION}
\label{sec:concl}

The cosmic rays (CRs) passing through the dark matter (DM) spike surrounding active galactic nuclei (AGN) can be cooled down due to their interactions with DM particles. In this work, assuming various $U(1)$ gauge symmetry models of the CRs-DM collisions, we explored the DM-proton and DM-electron scattering from CRs cooling in the AGN.

 We found that lowering the DM mass or mediator mass enhances the cooling rate compared to the standard model cooling rate. As a consequence, the cooling effect of CRs in AGNs is mostly important for low DM masses, specifically in the sub-GeV mass range. Further, using the CR cooling constraints in AGNs (NGC 1068 and TXS 0506+056), we limit the DM-electron and DM-proton couplings for both annihilating and non-annihilating DM particles. We found that for heavy mediator case, the electron cooling in TXS 0506+056 puts stringent constraints on low DM masses, i.e., $m_{\chi}\leq 2\times 10^{-4}$ GeV for non-annihilating, and $m_{\chi}\leq 2\times 10^{-5}$ GeV for annihilating DM particles. In the case of a light mediator, the constraints are weaker and limited by existing HB star constraints. Furthermore, we also reported that NGC 1068 provides a strong constraint on DM-proton elastic cross-section for the heavy mediator case. The bounds are stronger for $m_{\chi}\leq 10^{-2}$ GeV using NGC 1068, and $m_{\chi}\leq 2\times 10^{-4}$ GeV using TXS 0506+056. 

We emphasize that the CR cooling in AGNs provides stringent constraints on the DM-electron and DM-proton elastic cross-sections, which is stronger than the current direct detection and boosted DM constraints. Therefore, our analysis provides a novel parameter space for investigating the DM particles in the sub-GeV mass range. However, there is an uncertainty in the CR cooling timescales, which may influence the constraint on DM particles. Furthermore, upcoming observations will enable accurate modeling of the source and reduce the uncertainty, which will lead to robust constraints on the DM-electron and DM-proton cross-sections.  
 \section{ACKNOWLEDGEMENT}
This work is supported by the National Natural Science Foundation of China (NNSFC) under grants No. 12275134, No. 12335005, and No. 12147228 and the Special funds for postdoctoral overseas recruitment, Ministry of Education of China. AKM would like to thank Dr. Atanu Guha and Dr. Gonzalo Herrera for the fruitful discussions.

\bibliographystyle{utphys}
\bibliography{CRcooling}
\end{document}